# Markov chain Monte Carlo methods for hierarchical clustering of dynamic causal models

**Authors:**

Yu Yao[1], Klaas E. Stephan[1,2]

[1] Translational Neuromodeling Unit (TNU), Institute for Biomedical Engineering, University of Zurich & ETH Zurich, Switzerland.

[2] Max Planck Institute for Metabolism Research, Cologne, Germany.

**Abstract:**

In this paper, we address technical difficulties that arise when applying Markov chain Monte Carlo (MCMC) to hierarchical models designed to perform clustering in the space of latent parameters of subject-wise generative models. Specifically, we focus on the case where the subject-wise generative model is a dynamic causal model (DCM) for fMRI and clusters are defined in terms of effective brain connectivity. While an attractive approach for detecting mechanistically interpretable subgroups in heterogeneous populations, inverting such a hierarchical model represents a particularly challenging case, since DCM is often characterized by high posterior correlations between its parameters. In this context, standard MCMC schemes exhibit poor performance and extremely slow convergence. In this paper, we investigate the properties of hierarchical clustering which lead to the observed failure of standard MCMC schemes and propose a solution designed to improve convergence but preserve computational complexity. Specifically, we introduce a class of proposal distributions which aims to capture the interdependencies between the parameters of the clustering and subject-wise generative models and helps to reduce random walk behaviour of the MCMC scheme. Critically, these proposal distributions only introduce a single hyperparameter that needs to be tuned to achieve good performance. For validation, we apply our proposed solution to synthetic and real-world datasets and also compare it, in terms of computational complexity and performance, to Hamiltonian Monte Carlo (HMC), a state-of-the-art Monte Carlo. Our results indicate that, for the specific application domain considered here, our proposed solution shows good convergence performance and superior runtime compared to HMC.

# 1  Introduction

Dealing with the heterogeneity in clinical populations represents an important challenge for neuroimaging. This is particularly the case for psychiatry where contemporary diagnostic classifications group together patients with presumably heterogeneous disease mechanisms (Owen, 2014; Stephan et al., 2016). This heterogeneity is one possible reason for the low success rate of clinical trials, and stratification (e.g. by clustering the population into specific subgroups) might considerably increase the power of clinical trials (Schumann et al., 2014). This is a particularly promising approach when such clusters or subgroups are not defined in terms of abstract data features, but are interpretable in terms of disease-relevant mechanisms (Stephan et al., 2016).

This technical note addresses the specific problem of applying Markov chain Monte Carlo (MCMC) to hierarchical clustering in the context of generative embedding (GE). GE refers to a mapping from data to feature space that is instantiated by a generative model (Brodersen et al., 2011). Put simply, GE boils down to using (a function of) posterior densities of model parameters in order to define a feature space for subsequent machine learning. By achieving theory-led dimensionality reduction jointly with interpretability of features (in terms of data-generating mechanisms embodied by the generative model), GE can both enhance the performance and interpretability of machine learning when applied to neuroimaging or behavioural data (for reviews, see Frässle et al., 2018; Stephan et al., 2017). However, the necessity of model inversion can make GE technically challenging, particularly in hierarchical settings.

Here, we deal with the hierarchical unsupervised case of GE, i.e., group-level clustering in a high-dimensional space of latent variables. More specifically, this paper deals with the challenge of inverting a hierarchically structured generative model that distinguishes clusters of latent parameters from other (subject-wise) generative models, the (equally latent) dynamics of processes governed by these parameters, and the observations resulting from these processes. Specifically, we focus on HUGE (Yao et al., 2018) where the cluster formulation is based on Gaussian mixture models and the subject-wise generative model is a dynamic causal model (DCM). DCM is a nonlinear dynamic system model for estimating effective (directed) brain connectivity from fMRI (Friston, Harrison, & Penny, 2003) or EEG/MEG data (David et al., 2006). Like almost any other biological dynamic system model (cf. Gutenkunst et al., 2007), it may exhibit high posterior correlations among some of its parameters (Stephan, Weiskopf, Drysdale, Robinson, & Friston, 2007), rendering model inversion a difficult task.

Such challenges associated with parameter estimation are not unique to dynamic system models of neuroimaging data. Generally, dynamic system models are popular in scientific areas – including systems biology, medicine and neuroscience – that require an understanding of complex data in terms of latent parameters that govern the evolution of observed timeseries. Their application has been aided by the development of a variety of model inversion methods based on Hamiltonian Monte Carlo (HMC, Calderhead & Girolami, 2011; Kramer, Calderhead, & Radde, 2014), MCMC (Xun, Cao, Mallick, Maity, & Carroll, 2013), variational inference (VI, Friston, Mattout, Trujillo-Barreto, Ashburner, & Penny, 2007; Meeds, Roeder, Grant, Phillips, & Dalchau, 2019), or gradient matching techniques (Calderhead, Girolami, & Lawrence, 2009; Wenk et al., 2019).

However, incorporating a dynamic systems model, such as DCM, into a hierarchical clustering model exacerbates the difficulties associated with model inversion due to the interaction between the estimation of the parameters of the DCM and the clustering model. In particular, standard MCMC methods display a tendency to fail to converge under these circumstances, an issue we address in this paper. The contributions of this paper are as follows. First, we identify key features in the structure of the hierarchical clustering model which contribute to the convergence issues observed with MCMC. Based on these insights, we then propose a heuristic solution tailored to hierarchical clustering, which aims to improve convergence while preserving computational complexity. We demonstrate the effectiveness of our solution on synthetic and real-world examples based on a hierarchical clustering model for DCM known as hierarchical unsupervised generative embedding (HUGE, Yao et al., 2018). Finally, we discuss the complexity and performance of our proposed solution in comparison to Hamiltonian Monte Carlo (HMC), an advanced, general purpose Monte Carlo method designed to solve the convergence issues of standard MCMC methods without relying on detailed knowledge of the specific application.

Our work significantly goes beyond previous work on parameter estimation for hierarchically structured generative models. For example, Raman, Deserno, Schlagenhauf, and Stephan (2016) used standard MCMC in an early version of the HUGE model, but did not provide a detailed analysis on speed of convergence or computational complexity. The same model is also discussed in Yao et al. (2018), who applied VI instead of MCMC. Despite being extremely efficient, VI suffers from a number of drawbacks, making the availability of a complementary MCMC-based inversion scheme desirable. Specifically, VI traditionally requires the use of conjugate priors, which may restrict the expressiveness of the model and makes model extensions difficult. In addition, the simplifying assumptions on the approximate posterior required for VI mean that VI lacks both the asymptotic exactness and the ability to approximate multi-modal posteriors afforded by MCMC. Other hierarchically structured generative models of dynamic systems include the parametric empirical Bayesian variant of DCM (Friston et al., 2016) and hierarchically structured dynamic system models for applications in system biology (Meeds et al., 2019). However, in the latter two models, assignments of data points to groups are not inferred, but have to be supplied with the data; these models can therefore only be used for supervised learning (see also Ahn, Haines, & Zhang, 2017).

The application of HMC to hierarchical models in general has been discussed in Betancourt and Girolami (2015). In the present paper, we focus on the combination of a hierarchical model structure with a dynamic systems model. In addition, we take a different approach to model inversion by attempting to augment standard MCMC with specialized proposal distributions. In this context, we will provide a detailed comparison of our proposed approach with HMC.

## 2 Methods

### 2.1 Markov chain Monte Carlo for Hierarchical Clustering

In hierarchical clustering, a group-level clustering model (such as a Gaussian mixture model in the case of HUGE) is combined with a subject-wise generative model in such a way that the generative model is used to fit the subject-specific data points, while the clustering model is used to cluster the estimates of the latent parameters of the subject-wise generative models.

Figure 1 shows the graphical model of this hierarchical clustering problem with $K$ clusters and $N$ subjects, which contains the HUGE model as a special case. For each of the $n = 1, \ldots, N$ subjects, $\theta_n$ represents the parameters of the generative model that connects the subject-specific data $y_n$ with the clustering model. Here, the clustering model is represented by a Gaussian mixture model, where cluster number k ($1 \leq k \leq K$) is described by its mean $\mu_k$ and log-precision $\kappa_k$ and is assigned a cluster weight $\pi[k]$. $\lambda_n$ denotes the precision of observation noise. Note that despite having a fixed number of clusters K, the model can accommodate clustering solutions with less than K cluster, by leaving some of the clusters empty. Hence, K should be viewed as an upper limit on the number of clusters expected in the dataset. This stands in contrast to traditional clustering methods like k-means.

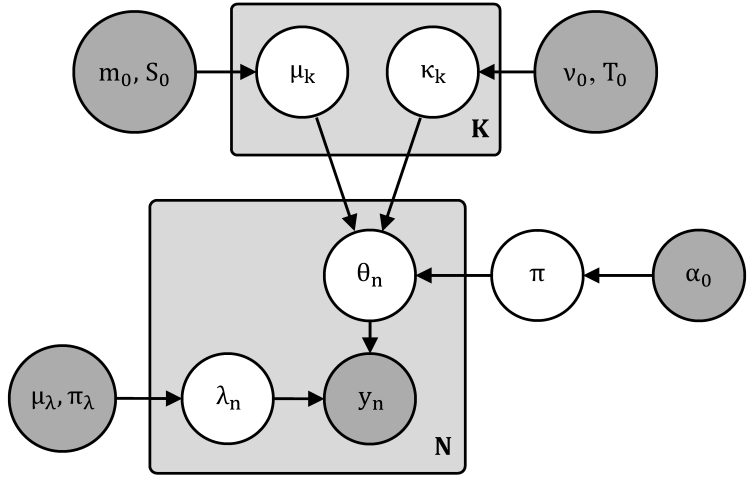

*Figure 1: Graphical model for the hierarchical clustering problem.*

Assuming that the observation model can be expressed as a, possibly nonlinear, transformation $g(\cdot)$ with additive Gaussian noise:

$$y_n = g(\theta_n) + \epsilon, \text{ with } \epsilon \sim N(0, exp(-\lambda_n)), \quad (1)$$

we can express the joint distribution of the hierarchical clustering model as follows:

$$p(y_n, \theta_n, \lambda_n, \mu_k, \kappa_k, \pi) = D(\pi|\alpha_0) \prod_{k=1}^{K} \{N(\mu_k|m_0, S_0) N(\kappa_k|v_0, T_0^{-1})\} \times$$

$$\prod_{n=1}^{N} \left\{ N(y_n|g(\theta_n), exp(-\lambda_n)) \, N(\lambda_n|\mu_\lambda, \pi_\lambda^{-1}) \sum_{k=1}^{K} \pi[k] N(\theta_n|\mu_k, exp(-\kappa_k)) \right\}, \quad (2)$$

where the symbols $N(\cdot)$ and $D(\cdot)$ denote the multivariate Normal and Dirichlet distributions, respectively. In addition, $\alpha_0$ denotes the parameter of the prior over cluster weights, $m_0$ and $S_0$ the prior mean and covariance of the cluster centres and $v_0$ and $T_0$ the prior mean and covariance of the cluster log-precision.

Performing Bayesian inference on this model requires the joint estimation of the parameters of the subject-wise generative models ($\theta_n$ and $\lambda_n$) and the parameters of the group-level clustering model ($\mu_k$, $\kappa_k$ and $\pi$). In principle, an attractive way to do this is to apply Monte Carlo sampling, which offers a number of advantages including asymptotic exactness, lack of conjugacy requirements, and the availability of well-established standard algorithms and software tools. One of these standard algorithms is Metropolized Gibbs sampling, which can be applied almost universally to any target distribution which can be evaluated up to a multiplicative constant for arbitrary parameter combinations (Gelman, 2014). It works by sampling parameters, or groups of parameters, in turn from their conditional distribution given the remaining parameters of the model, employing the Metropolis-Hastings (MH) algorithm whenever it is not possible to sample from one of the conditional distributions directly. Specifically, for hierarchical clustering, this means sampling each of the following parameters $\theta_n, \lambda_n, \mu_k, \kappa_k$ and $\pi$ ($1 \leq n \leq N, 1 \leq k \leq K$) from its conditional distribution given all the remaining parameters, which is done using MH, due to the lack of conjugacy. The exact forms of these conditional distributions are given in the Supplementary Material.

However, a major weakness of Gibbs sampling in particular, and MH in general, is slow convergence in the case of highly correlated parameters (Bishop, 2006). In our case, the structure of hierarchical clustering induces a strong correlation between the parameters of the clustering model, specifically $\mu_k$, and the parameters of the generative model $\theta_n$. This can be understood by noting that $\mu_k$ is the parent of $\theta_n$ in the graphical model in Figure 1. Additionally, the parameters $\mu_k$, $\kappa_k$ and $\pi$ are strongly anti-correlated, as they are co-parents of $\theta_n$. Note that despite being nominally unobserved, we are conditioning on the current sample value of $\theta_n$ when drawing the parents during Gibbs sampling. Hence, these variables suffer from the well-known "explaining away" effect (Bishop, 2006).

When applying Gibbs sampling to hierarchical clustering in practice, these issues lead to very characteristic failure modes where either some data points are stuck in the wrong cluster, or an empty (or almost empty) cluster is stuck close to the prior mean $m_0$. These issues are exacerbated by generative models $g(\theta_n)$ with high posterior correlations among their parameters; an issue that is frequently found in generative models involving dynamical system formulations, such as DCM. Examples illustrating these failure modes are provided in section 3.

In order to address convergence issues, advanced sampling methods have been developed. Hamiltonian Monte Carlo (HMC, Betancourt & Girolami, 2015; Calderhead & Girolami, 2009; Duane, Kennedy, Pendleton, & Roweth, 1987) is generally considered to be the state-of-the-art in this field, designed to increase sampling efficiency and speed up convergence for a wide range of target distributions. However, this generality comes at the cost of higher complexity, both numerically and in terms of implementation effort. In addition, it introduces the need to tune additional hyperparameters to achieve optimal performance (Behrens, Friel, & Hurn, 2012; Betancourt, 2016). In the following section, we introduce an alternative solution with the goal of improving convergence of the Metropolized Gibbs sampler for hierarchical clustering, while minimizing the additional computational complexity.

## 2.2 Improving Convergence for Hierarchical Clustering

In the previous section, we have identified key features in hierarchically structured clustering models which are responsible for convergence issues in MCMC-based inversion of these models. In order to address these issues, we suggest constructing specialized proposal distributions tailored to the dependency structure of the hierarchical clustering model.

Specifically, we suggest using special proposal distributions during the MH phase of the Metropolized Gibbs sampler which depend on the sample value of the parameters which are not being sampled at the current step. For example, when sampling from the conditional distribution over cluster means $\mu_k$ given all remaining parameters, one may use a proposal distribution which depends on the current sample value of the other parameters (in the following, the $(\tau)$ in the exponent denotes the current sample value, while the star in the exponent denotes the proposal value):

$$q(\mu_k^*) = q(\mu_k^* | \theta_n^{(\tau)}, \pi^{(\tau)}, \ldots), \tag{3}$$

This departs from standard proposal distributions, such as using a Gaussian kernel centred on the last sample value of $\mu_k$ itself:

$$q(\mu_k^*) = N\left(\mu_k^* \middle| \mu_k^{(\tau)}, \sigma_{MH}\right). \tag{4}$$

Note that our idea does not violate detailed balance since the target distribution is the conditional distribution, and we only use the sample values of the parameters we are currently conditioning on.

In theory the optimal choice for such a proposal distribution would be the conditional distribution itself. However, the inability to sample from the conditional distribution directly is what necessitated the MH step in the first place. In practice, we therefore alternate between (i) a standard proposal distribution, like a Gaussian kernel centred on the last sample, which is used to explore the current posterior mode, and (ii) a special proposal distribution, designed to disrupt the random walk behaviour of the chain, for example, by proposing jumps to possible locations of other modes of the posterior. The key is that these special proposal distributions should be more effective if they are informed by the current sample value of the other parameters. Note that alternating between different transition operators in MH is valid, as long as each individual transition operator is valid (Brooks, Gelman, Jones, & Meng, 2011).

The idea of disrupting random walk behaviour in MH using specialized proposal distributions – for example derived from extensive domain knowledge – is not new. In fact, our approach was inspired by Carlin and Chib (1995) who, after reformulating a model selection problem in terms of a clustering model, faced convergence issues similar to those seen in hierarchical clustering. However, what distinguishes our approach from previous methods is the insight that, in the case of hierarchical clustering, the special proposal distributions can make use of the sample value of the parameters currently not being sampled to identify promising proposals. For example, when sampling from the conditional distribution over clustering parameters, one may use a proposal distribution informed by the current sample value of the subject-level parameters. This eliminates the need for designing proposal distributions based on domain knowledge, making the method less application dependent. In addition, it also eliminates the need for tuning the special proposal distribution in preliminary test runs of the sampler, as was done by Carlin and Chib (1995).

Based on the issues with hierarchical clustering identified in the previous section, we focus on two steps in the Metropolized Gibbs sampler: (i) sampling from the conditional distribution over $\theta_n$ and (ii) sampling from the conditional distribution over the cluster parameters $\pi$, $\mu_k$ and $\kappa_k$.

For step (i), our special proposal distribution is extremely simple: sample the proposal $\theta_n^*$ randomly from the clustering model:

$$q(\theta_n^*) = \sum_{k=1}^{K} \pi^{(\tau)}[k] N\left(\theta_n^* \big| \mu_k^{(\tau)}, \exp(-\kappa_k^{(\tau)})\right), \qquad (5)$$

In order to satisfy detailed balance, we need to derive the corresponding MH acceptance rate, which is given by:

$$a = \min\left\{1, \frac{p(\theta_n^*) q(\theta_n^{(\tau)} | \pi^{(\tau)}, \mu_k^{(\tau)}, \kappa_k^{(\tau)})}{p(\theta_n^{(\tau)}) q(\theta_n^* | \pi^{(\tau)}, \mu_k^{(\tau)}, \kappa_k^{(\tau)})}\right\}. \qquad (6)$$

Inserting the expressions for the conditional distribution (see Eq. (S5) in the Supplementary Material) and the proposal distribution from Eq. (5), the acceptance ratio simplifies to the ratio of likelihoods, since all terms depending on the cluster parameters cancel out:

$$a = \min\left\{1, \frac{N\left(y_n \big| g(\theta_n^*), \exp(-\lambda_n^{(\tau)})\right)}{N\left(y_n \big| g(\theta_n^{(\tau)}), \exp(-\lambda_n)\right)}\right\}. \qquad (7)$$

Therefore, this kernel has the convenient property that it has no free hyperparameters which need to be tuned. In fact, the only parameter that needs to be tuned in the entire approach is the frequency with which to propose from this distribution as compared to the standard Gaussian kernel. In the experiments presented in section 3, this frequency was chosen during a preliminary test run of the sampler and kept fixed for all subsequent experiments. This was done in order to keep the setup as simple as possible. However, the tuning process can in principle be accomplished during the burn-in phase of the sampler, removing the need for any preliminary test runs.

The second special proposal density (step ii above) was designed to address the problem of transitioning between different posterior modes. Specifically, this proposal jointly samples cluster model parameters $\pi$, $\mu_k$ and $\kappa_k$ given the current sample values of $\theta_n$, while allowing transitions between posterior modes associated with different number of clusters. This is achieved using the following process. First, we sample a number k uniformly between 1 and K, where K is the maximum number of clusters in the model. Then, we use k-means to identify a plausible clustering solution with exactly k clusters and draw the parameters, i.e. weight, mean and covariance, of the first k clusters from a distribution centred on this solution. Finally, if k is smaller than K, the parameters of the remaining clusters from k+1 to K are sampled from the prior over clusters. The last step ensures that the overall number of parameters stays constant, avoiding the complications arising from changing the dimensionality of the parameter space (Green, 1995).

Given that this proposal density requires the use of k-means to obtain an intermediate clustering solution, it may appear that considerable computational overhead is being introduced. However, our experiments in section 3 show that the computational cost introduced by our special proposal distributions is almost negligible. The derivation of the acceptance ratio for this proposal distribution can be found in the Supplementary Material. The key feature of this proposal is that it allows transitions between posterior modes representing clustering solutions with different number of clusters. For example, in certain situations, the solution where all data points belong to one big cluster might be equally plausible as the solution where data points are divided into two smaller clusters. Transitioning between these two solutions is extremely difficult for standard Gibbs sampling, but is possible with the proposal density introduced above.

Also, note that the proposal introduced above makes a joint update to cluster weights, means and covariances. This is in contrast to standard Gibbs sampling, where these variables are sampled successively from their respective conditional distributions. This makes the standard Gibbs sampling susceptible to the strong posterior correlations between these parameters.

In section 3, we present empirical evidence that our proposed approach significantly improves convergence for hierarchical clustering while avoiding the added complexity of advanced Monte Carlo schemes. To this end, we apply our method to a hierarchical clustering model built on a subject-specific generative model of effective brain connectivity; the latter, a dynamic causal model (DCM) of fMRI data is briefly introduced in the next section.

## 2.3 Dynamic Causal Modelling

The approach introduced in section 2.2 applies to hierarchical clustering, irrespective of the subject-specific generative model. However, for the remainder of this paper, we will focus on a hierarchical clustering model known as hierarchical unsupervised generative embedding (HUGE, Yao et al., 2018), which combines clustering with a class of dynamic systems models called dynamic causal modelling (DCM). In the following, we provide a short introduction to DCM; specifically, its implementation for functional magnetic resonance imaging (fMRI) data. For a more detailed description of DCM for fMRI, we refer to the original paper by Friston et al. (2003).

DCM is a class of generative models for inferring effective connectivity between brain regions from fMRI data (Friston et al., 2003) or electro- or magnetoencephalography data (David et al., 2006). A DCM for fMRI consists of an evolution function (formulated as a bilinear or non-linear dynamic systems model) which is linked to a nonlinear observation function. Specifically, the evolution function $f(\cdot)$ describes the temporal evolution of neuronal population states $x$ in a network of brain regions, together with the evolution of activity-induced hemodynamic states $s$:

$$(\dot{x}, \dot{s})^T = f(x(t), s(t), u(t)) \qquad (8)$$

The evolution function is parameterised by parameters $\theta$ (see below) and under the influence of known perturbations or external inputs $u$. These inputs $u$ could represent, for example, sensory stimuli or cognitive interventions (such as cued attention).

Neuronal and hemodynamic states are linked via a nonlinear observation function $h(\cdot)$ to the observed fMRI data $y(t) = h(x(t), s(t)) + e(t)$, under Gaussian assumptions about the noise (and dealing with non-IID properties). For simplicity, this description ignores region-specific parameters of this observation function; for more details on the hemodynamic model, we refer to Friston, Mechelli, Turner, and Price (2000) and Stephan et al. (2007).

Of particular interest for practical applications is the parameterization of the neuronal evolution function, i.e. the part of $f(\cdot)$ which describes the evolution of the neuronal activity $x$, consisting of a bilinear evolution function with parameter matrices $A$, $C$ and the set $B = \{B^{(l)}: l = 1, \dots, L\}$ containing one matrix per input:

$$\dot{x} = Ax + \sum_{l=1}^{L} u_l B^{(l)} x + Cu \qquad (9)$$

Here, $A$ represents the endogenous connectivity, i.e. the connectivity between regions in the absence of external influences. $B^{(l)}$ represents the modulatory influence of input $l$ on the endogenous connectivity in $A$. And finally, $C$ represents the strength of inputs that drive the regions directly.

Evaluation of DCM as a generative model requires numerical integration of the evolution function $f$, which, assuming the use of an efficient integrator like the Euler method, requires $O(TLR^2)$ operations, where $R$ denotes the number of regions and $T$ the length of the fMRI time series. In comparison, the complexity of evaluating the observation function $h$ is negligible. Note also that in contrast to other domains which apply dynamic system models (Kramer et al., 2014), the process of fitting DCM to neuroimaging data is mostly driven by transients and less by the steady-state.

When using DCM as the subject-wise generative model in hierarchical clustering, the model parameters of interest would be represented by $\theta = \{A, B, C\}$, while the predicted fMRI time series result from the process of first integrating the DCM evolution equation $f(\cdot)$ and then applying the observation function $h(\cdot)$.

## 2.4 Hamiltonian Monte Carlo

Hamiltonian Monte Carlo (HMC) is considered to be the state-of-the-art in the field of general purpose Monte Carlo sampler capable of avoiding random walk behaviour and obtaining less correlated samples. However, the efficiency of HMC comes at the cost of increased computational complexity per sample, which may offset the benefit of being able to use shorter chains. This is because in order to obtain a new sample, HMC needs to simulate Hamiltonian dynamics in a potential landscape defined by the target distribution, which requires numerically integrating a dynamical system (Brooks et al., 2011). In this section, we present a theoretical analysis of the complexity of HMC for hierarchical clustering models of DCM in comparison with more conventional sampling techniques.

Theoretical analysis shows that the complexity of HMC of $O(D^{5/4})$ compares favourably to the complexity of basic methods, such as Gibbs sampling, of $O(D^2)$ (Hoffman & Gelman, 2014), where $D$ denotes the dimensionality of the parameter space. However, this complexity refers to the number of samples needed to explore the target distribution and does not account for the complexity of obtaining each sample. On a sample-by-sample level, HMC introduces two sources of additional complexity: the evaluation of gradients of the target distribution and the simulation of the Hamiltonian dynamics. While gradient evaluation may be addressed with automatic differentiation techniques without increasing the order of numerical complexity

(Baydin, Pearlmutter, Radul, & Siskind, 2018), simulating the Hamiltonian dynamics requires a symplectic integration scheme such as the leapfrog integrator (Brooks et al., 2011), which needs two evaluations of the gradient of the target density per integration step.

Unfortunately, in a dynamic systems model, such as DCM, evaluation of the joint distribution and the associated gradients is often the most expensive part. For DCM, the complexity of a single evaluation is given by $O(TLR^2)$. Hence, evaluating the joint distribution for a DCM-based hierarchical clustering model, such as HUGE, with data from $N$ subjects would require on the order of $O(TNLR^2)$ operations due to the DCM part of the model alone. At the same time, the number of parameters of such a model is approximately given by $D = (N + K)LR^2 \approx NLR^2$ (assuming $N \gg K$). This means that the evaluation of the joint distribution alone would contribute $O(TD)$ and $O(JTD)$ operations to Gibbs sampling and HMC, respectively, where J denotes the number of steps used by the leapfrog integrator in HMC.

The optimum value for $J$ depends on the structure of the target distribution itself and is difficult to determine a priori (Betancourt, 2016). However, for typical DCMs, the length of the fMRI time series $T$ is approximately on the same order as $LR^2$ (Friston et al., 2003). Hence, it becomes clear that the cost of evaluating the DCM could very well dominate the complexity of the entire inference algorithm, and may even negate the advantage afforded by HMC in terms of providing more independent samples if $JT$ is of the same magnitude as $D$.

In the next section, we investigate this problem from an empirical perspective by comparing the performance and computation time of HMC and Metropolized Gibbs sampling on a set of synthetic and real-world datasets.

## 3 Results

In this section, we present results from two experiments which illustrate the convergence issues encountered with standard Gibbs sampling for hierarchical clustering, as well as the improvements achieved using the approach proposed in section 2.2. For comparison, we also ran both experiments using HMC. For this purpose, we chose the No-U-Turn sampler (NUTS) provided by stan (Stan Development Team, 2020), because of its ability to automatically tune the hyperparameters of HMC for optimal performance. The code used to run these experiments will be made available as part of the open-source toolbox TAPAS (Translational Neuromodeling Unit, 2014). To account for issues related to label switching, samples were relabelled with the approach from Stephens (2000) throughout all our experiments. This method was chosen because it unifies several previously established relabelling schemes for MCMC. Details on the computing environment can be found in the Supplementary Material.

Since the goal of this technical report is the development of an efficient method for improving convergence of Gibbs sampling for hierarchal clustering, we focus our quantitative analysis on the assessment of the convergence of the samplers and the correctness of the inference for synthetic data with available ground truth. While convergence is assessed using the potential scale reduction factor (PSRF) proposed by Brooks and Gelman (1998), the quality of inference can be most conveniently summarised by calculating the balanced purity (Brodersen et al., 2014) of the clustering result. Although it may seem tempting to use a measure such as the root mean squared (RMS) error of the posterior mean of the DCM parameters, it should be noted that, for models with high posterior covariance between its parameters, such as DCM, the posterior may extend over a larger area or even be multi-modal, rendering the posterior mean less informative. In most circumstances, an RMS-based measure would likely reflect the posterior covariance of the model instead of the accuracy of the inversion scheme. In the following, we provide a short introduction to the PSRF and the balanced accuracy.

The PSRF was introduced by Brooks and Gelman (1998) in order to assess the convergence of an MCMC sampler. It quantifies the consistency between independent chains by comparing the variance of the samples within each chain with the variance of samples between chains. Given $m$ chains each containing $n$ samples the PSRF calculated via the formula:

$$\hat{r} = \frac{m+1}{m}\frac{\hat{\sigma}_+^2}{W} - \frac{n-1}{mn} \text{ with } \hat{\sigma}_+^2 = \frac{n-1}{n}W + \frac{B}{n}, \tag{10}$$

where $W$ and $B/n$ denote the within-chain variance and between-chain variance, respectively. Upon converged, the PSRF should approach a value of 1. In experimental settings, a value between 1 to 1.1 is generally accepted to indicate convergence of the sampler

The balanced purity (Brodersen et al., 2014) is a measure of the quality of clustering solutions, where a value of 1 indicates a perfect result, while a value of 0.5 indicates random assignment (for K=2). Given estimated $\Omega = (\omega_1, \ldots, \omega_K)$ and true $C = (c_1, \ldots, c_J)$ class assignments, the balanced purity is calculated from the purity using the formula:

$$bp(\Omega, C) = \left(1 - \frac{1}{K}\right)\left(\frac{purity(\Omega, C) - \xi}{1 - \xi}\right) + \frac{1}{K} \tag{11}$$

where $\xi$ is the degree of imbalance (i.e. the fraction of subject associated with the largest class) and the purity is defined as:

$$purity(\Omega, C) = \frac{1}{N}\sum_{k=1}^{K} \max_j |\omega_k \cap c_j|. \tag{12}$$

Here, $|\omega_k \cap c_j|$ denotes the number of subjects in cluster $k$ which are associated with the true class $j$. The advantage of using the balanced purity score is that it accounts for the biasing effects of an imbalanced dataset, which affects the usefulness of the more traditional purity score in dataset containing classes of varying sizes.

### 3.1 Experiment 1: Synthetic Data

In the first experiment, we used synthetic data with known ground truth to validate and compare the different inversion methods. For this purpose, we generated 20 synthetic datasets, where each dataset contained 30 simulated subjects divided into two clusters with different patterns of network connectivity. Repeating the experiment for 20 datasets introduces a range of variations which ensures that any observed differences between the samplers are not simply due to random properties of a particular dataset.

The time series data of each subject were generated using the 3-region DCM shown in Fig 2A with connectivity parameters drawn randomly from a Gaussian distribution centred on the respective cluster mean. The cluster means were chosen to represent the two distinct connectivity patterns shown in Fig 2B and 2C. The structure of this DCM was inspired by an actual experimental design from van Leeuwen, den Ouden, and Hagoort (2011). The numerical values of the ground truth parameters are listed in the Supplementary Material.

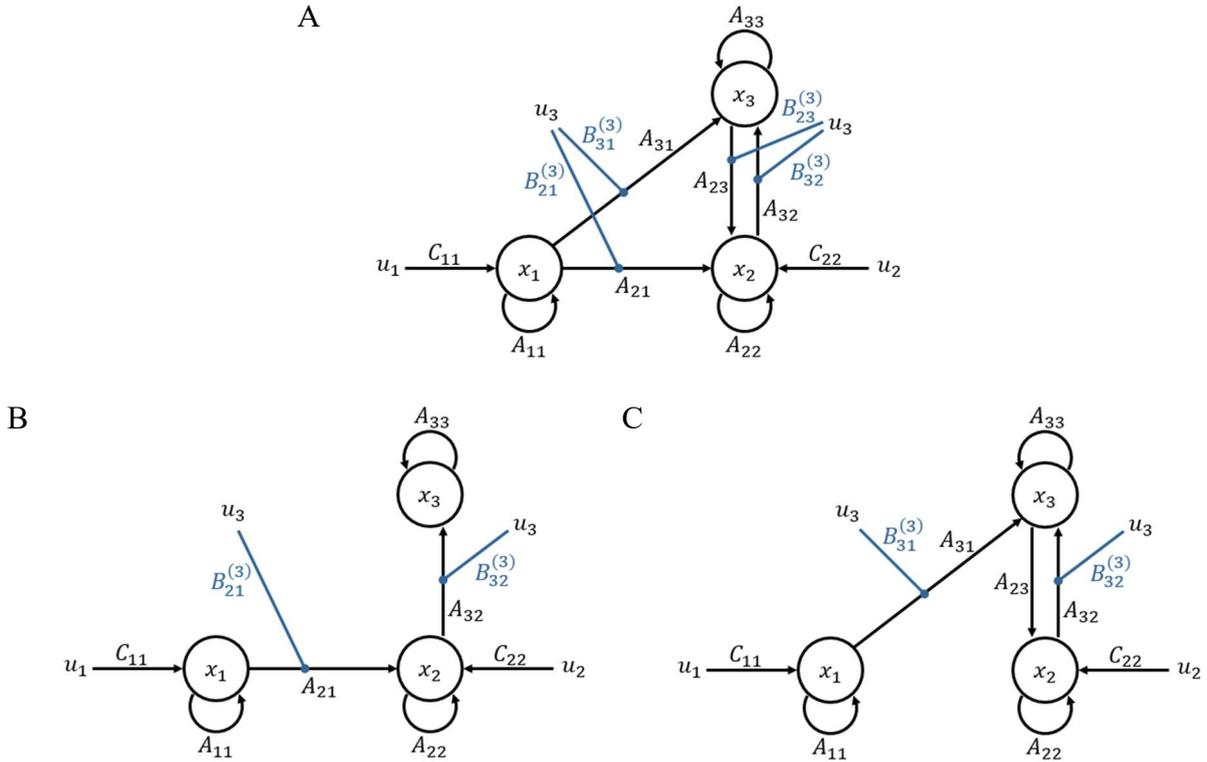

*Figure 2: A: DCM network for generating synthetic datasets. B: Sub-network corresponding to the first cluster. C: Sub-network corresponding to the second cluster.*

For each dataset, we inverted the HUGE model using 3 different approaches: standard Metropolized Gibbs sampling, the improved version described in section 2, and HMC as implemented in stan. For the Gibbs samplers, 4 independent chains with $1 \times 10^5$ samples each were run, which took less than 3.5 hours on average. The first half of each chain was discarded for burn-in before convergence was assessed using the PSRF.

Figure 3 (top panels) shows that PSRF values across all synthetic datasets for our improved Gibbs sampler are mostly within the range of 1 to 1.1 which is generally accepted to indicate convergence of the sampler. However, for the standard Gibbs sampler, the PSRF values indicate failure of convergence for some of the datasets. This impression is confirmed by Figure 4, showing histograms of the balanced purity over the datasets. In the comparison between standard and improved Gibbs sampler, we observe a clear improvement of the overall performance across the datasets.

In order to gain a better understanding of the failure mode of standard Gibbs, we chose one representative dataset and plotted the posterior estimates of subject assignment estimated for each chain individually (Figure 5). This reveals that for standard Gibbs the chains are stuck in different local maxima of the posterior density corresponding to different subject assignments. At the same time, Figure 6 shows a difference of over 100 in the mean log-joint probability between the chains. This indicates that some chains are most likely stuck in low probability regions of the parameter space which are hard to get out of.

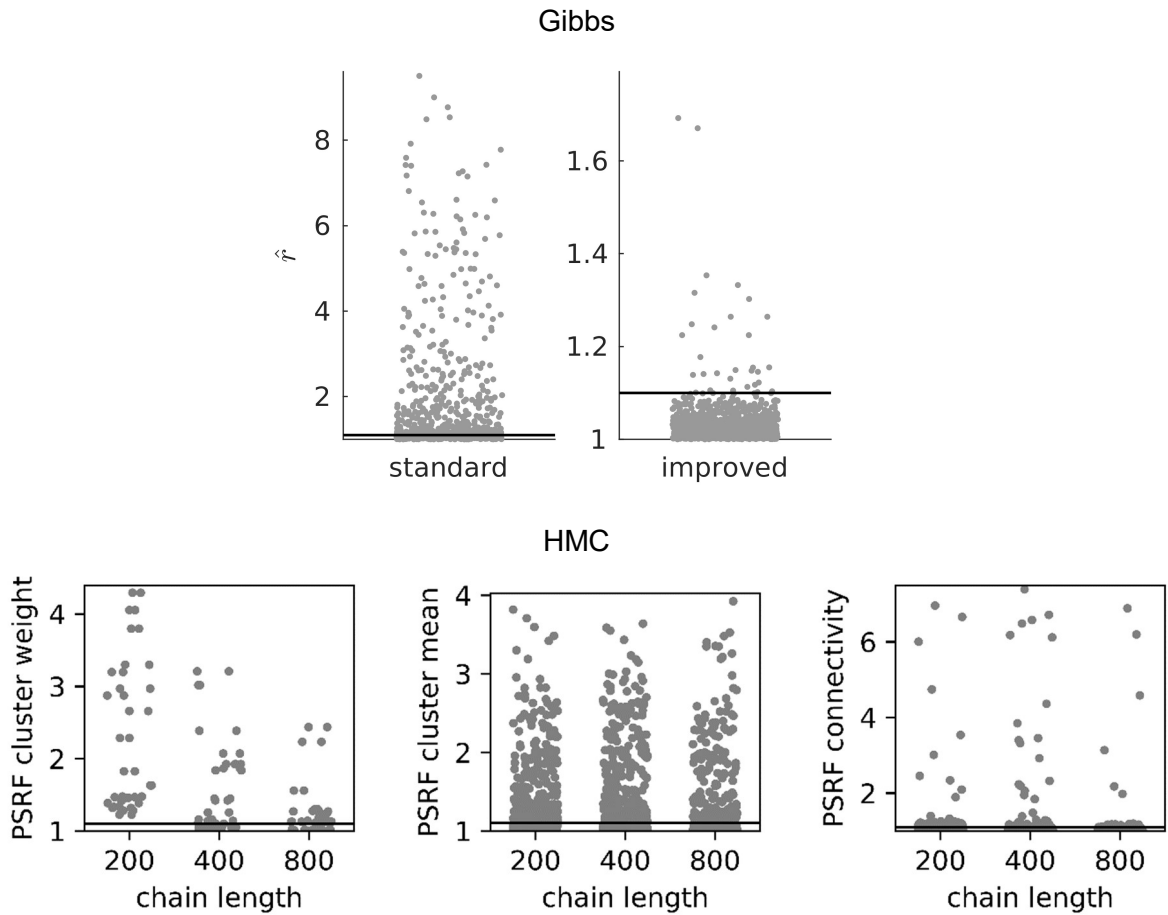

*Figure 3: Top: Range of PSRF of standard and improved Gibbs sampling observed across all synthetic datasets. Bottom: Range of PSRF values for different sets of parameters obtained with HMC for chains of length 200, 400 and 800 samples. Horizontal black lines mark the threshold of PSRF=1.1, which is commonly accepted to indicate convergence.*

On the other hand, for the improved version, all chains converged to the same area of the parameter space (Figure 5, right) with roughly the same mean log-joint probability (Figure 6, right). In addition, the estimated posterior subject assignment closely matches the ground truth assignment.

In order to compare our improved Gibbs sampler to HMC, we implemented the HUGE model in stan and inverted the model for each of the synthetic datasets. As with our Gibbs sampler, stan samples 4 independent chains per dataset, discards the first half of each chain for burn-in and pools the remaining samples over all chains to obtain convergence statistics and posterior quantiles. Since HMC has been designed for efficiency, HMC requires less samples to explore the target distribution than a sampler based on random-walk Monte Carlo. Hence, it is not appropriate to choose the same chain length for HMC and Gibbs sampling. Instead a more sensible approach would be to run HMC until convergence, as assessed by the PSRF, and compare how many sample and how much computation time were required relative to Gibbs sampling. In time sensitive applications, which might arise, for example, in clinical settings, an alternative approach would be to allocate a time budget and compare which sampler is able to converge within this budget. Following this strategy, we repeated the entire experiment three times with three different setting for the chain length: 200, 400 and 800 samples. Based on preliminary experiments, we predicted that these settings would limit the computation time to less than a day. For comparison, the Gibbs sampler converged within 3.5 hours on average.

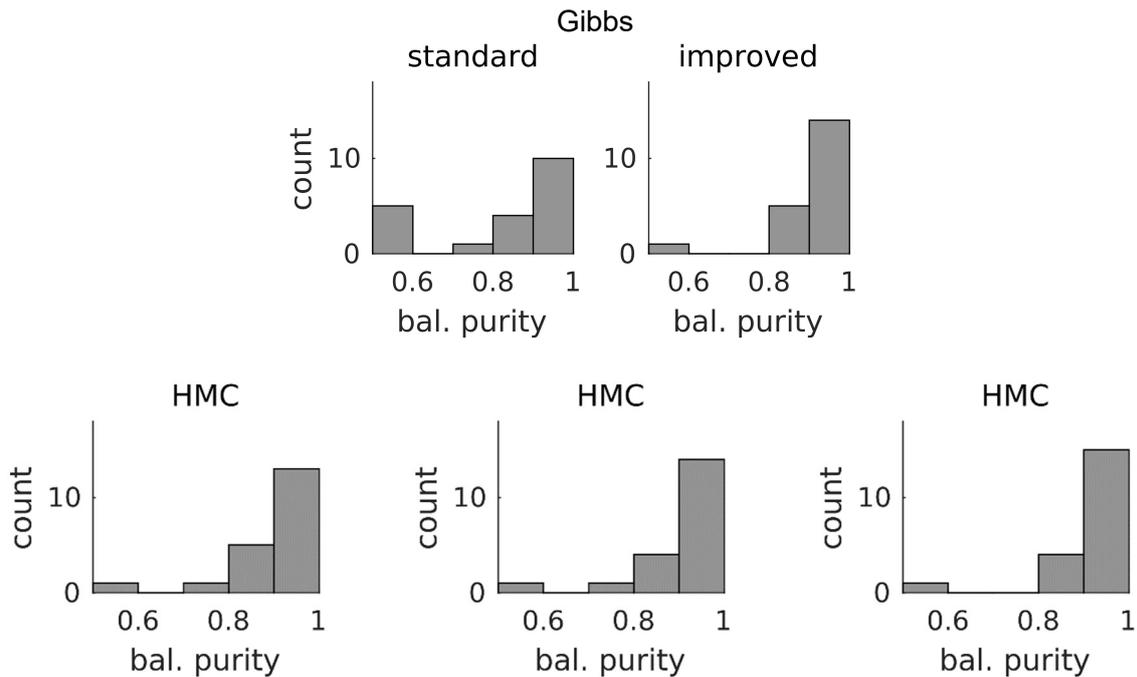

*Figure 4: Top: Histograms over balanced purity values obtained with standard (left) and improved (right) Gibbs sampling. Bottom: Histograms over balanced purity values obtained with HMC for chains of length 200 (left), 400 (middle) and 800 (right).*

The histograms over balanced purity values in Figure 4 show that, using chains of 800 samples, HMC achieved a similar clustering performance as our improved Gibbs sampler with $10^5$ samples, confirming a superior efficiency per sample for HMC. However, Figure 3 reveals that the PSRFs are far from the value of 1, especially for the cluster mean and DCM connectivity parameters, indicating that the sampler has not converged, even for the maximum setting of 800 samples.

A simple solution would be to increase the length of the chains. Unfortunately, Figure 7 shows that even for the shortest chains of 200 samples, the computation time of HMC exceeds that of Gibbs sampling with $10^5$ samples. Note that these experiments where conducted on a high performance computing (HPC) cluster equipped with processors which differ in speed. However, when repeating the sampling with HMC on a local workstation equipped with more powerful processors than the HPC cluster, we did not observe any significant speedup.

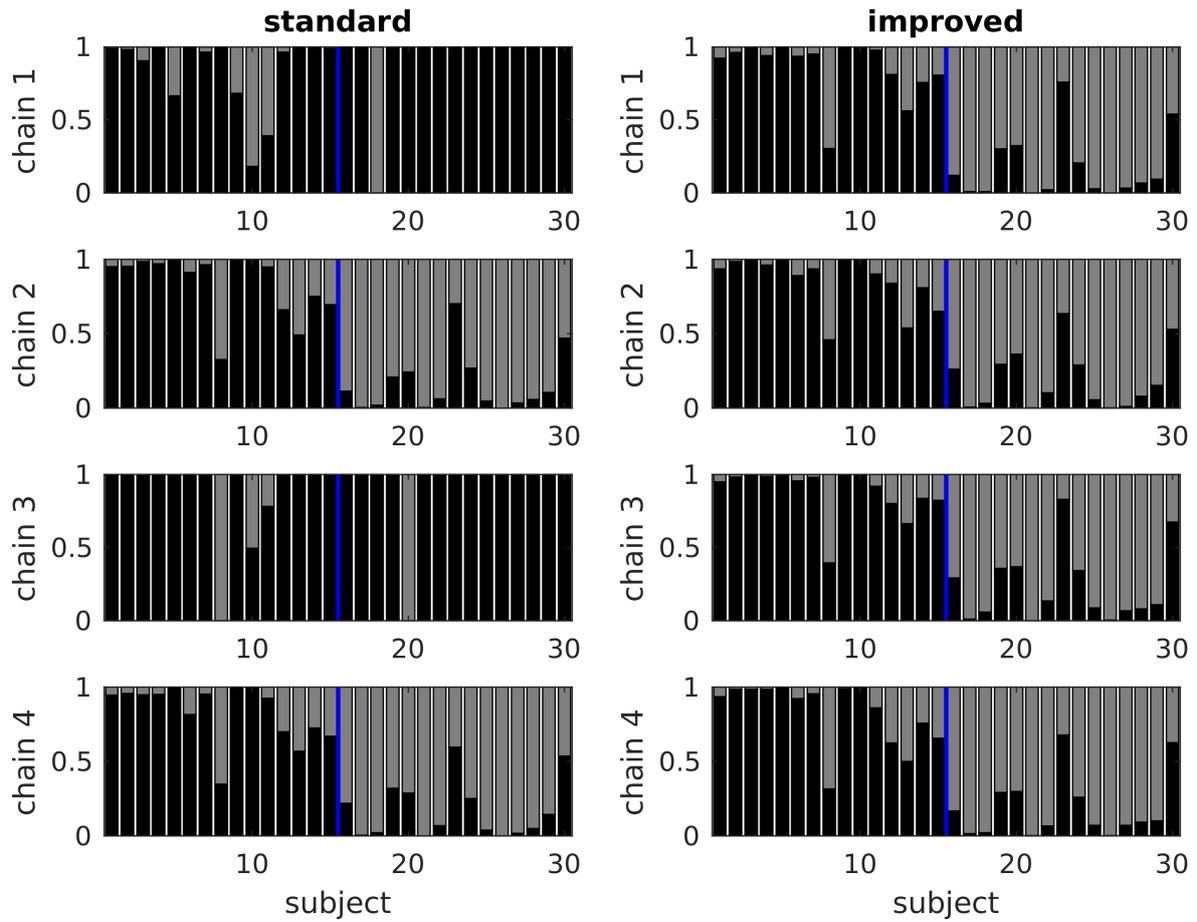

*Figure 5: Posterior subject assignment for one exemplary synthetic dataset, estimated for each chain individually with standard (left) and improved (right) Gibbs sampling. The blue line separates the first 15 subject generated from the first cluster from the last 15 subjects generated from the second cluster.*

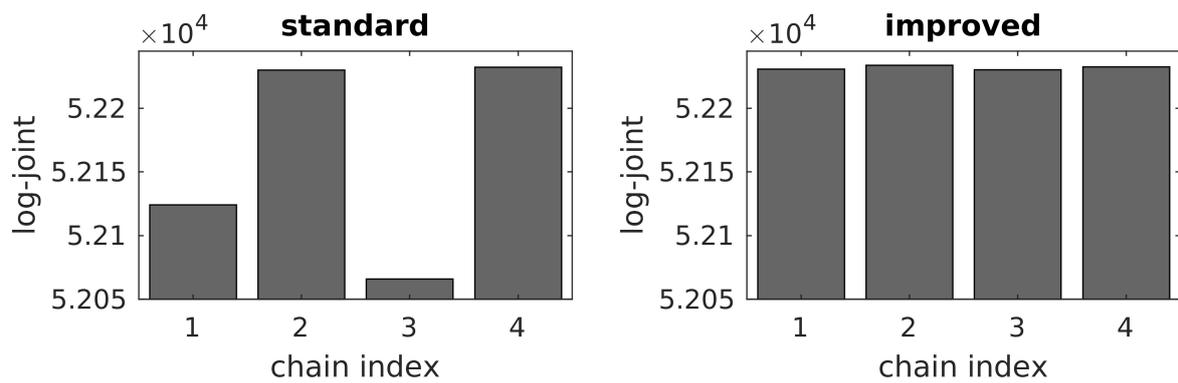

*Figure 6: Mean log-joint probability of individual chains for the synthetic dataset in Figure 5 obtained with standard (left) and improved (right) Gibbs sampling.*

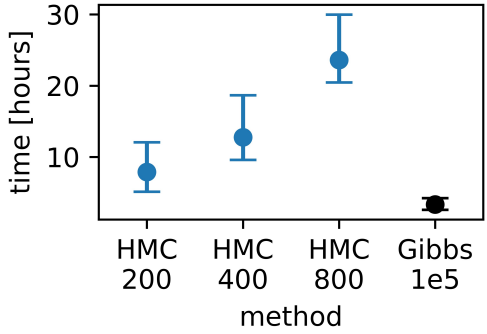

*Figure 7: Comparison of the range of computation times for HMC (blue) and Gibbs sampling (black). HMC200: HMC with 200 samples, HMC400: HMC with 400 samples, HMC 800: HMC with 800 samples and Gibbs1e5: Gibbs sampling with $10^5$ samples.*

## 3.2  Experiment 2: Experimental Dataset

In the second experiment, we analysed a real-world dataset from an fMRI experiment investigating speech perception in stroke patients compared to healthy controls. The cohort included 26 healthy controls and 11 patients. For details of the experimental setup, see Leff et al. (2008). In a previous DCM-based analysis of this dataset (Schofield et al., 2012), a task-relevant network containing 6 regions (with 3 regions in either hemisphere) was identified. For hierarchical clustering, we simplified the network by removing two subcortical regions with low signal-to-noise ratio, resulting in the 4-region network shown in Figure 8. This is in line with the approach from Yao et al. (2018).

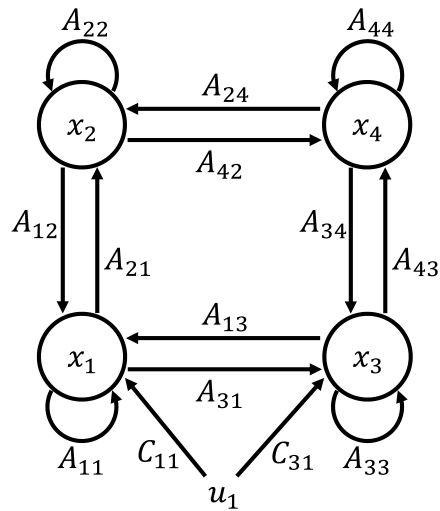

*Figure 8: The 4-region DCM network structure used for the analysis of the experimental dataset.*

As before, we inverted HUGE using both standard and improved Gibbs sampling and HMC. Settings were kept identical to the synthetic case except for the chain length, which was increased to 600,000 samples including 300,000 burn-in for Gibbs. The choice of chain length was partly based on previous experience with a similar clustering model (Raman et al., 2016). As before, individual chains were sampled in parallel on an HPC cluster, which required a computation time of $140 \times 10^3$ seconds (approx. 40 hours) on average for both versions of Gibbs sampling. Note that the computation time of the improved version of Gibbs is on average only

4% higher than that of standard Gibbs, indicating that the overhead introduced by our special proposal distributions is almost negligible. Also note that computation time is not a linear function of the number of regions of the DCM, but depends on many factors such as the number of connections, the number of subjects, the length of the time series, the type of the DCM (linear, bilinear, nonlinear), etc.

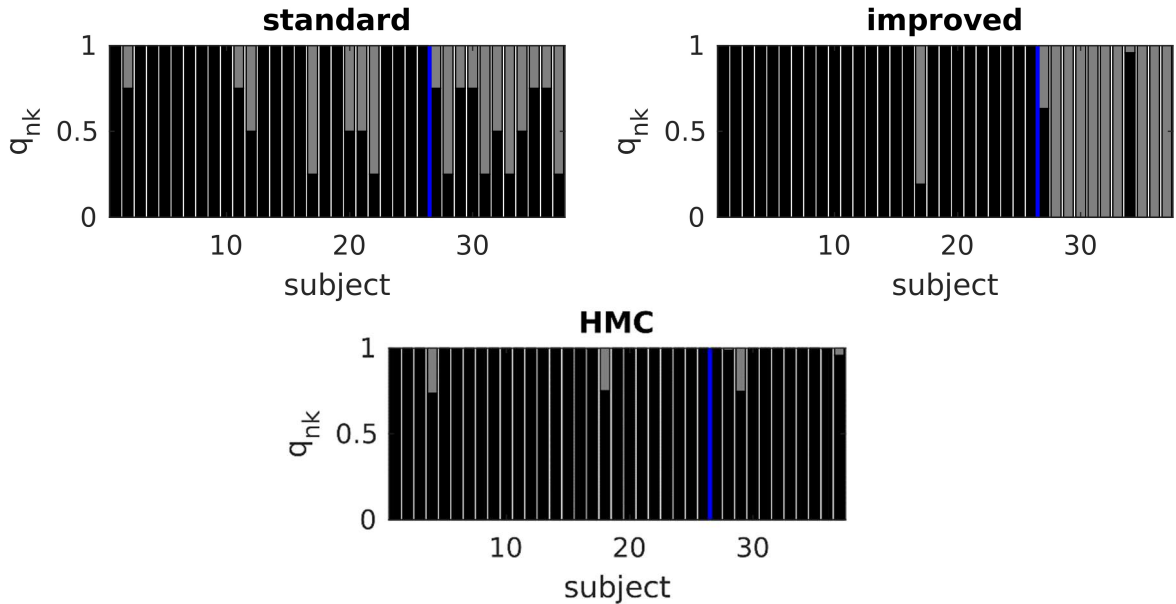

*Figure 9: Subject assignment for experiment 2, estimated from samples pooled over all chains. Top left: standard Gibbs, top right: improved Gibbs, bottom: HMC. The blue line separates the controls (first 26 subject) from patients (last 11 subjects).*

Figure 9 shows the posterior assignment estimated from samples of all 4 chains. It is apparent that the clusters obtained with improved Gibbs sampling match far better onto the known subgroups of controls and patients. The balanced purity achieved with the improved method is 86%, whereas standard Gibbs sampling only achieves 60%. The reason for this difference becomes apparent when plotting the posterior assignment estimated for each chain individually. Figure 10 shows that, similar to experiment 1, the chains from standard Gibbs are stuck in different local maxima of the posterior, while for the improved version, all chains converged to the same configuration. We again compared the mean log-joint probability between chains (Figure 11) and observed a difference of 100 for standard Gibbs and only 5 for the improved method.

This observation is also reflected in the PSRF (Figure 12, top), which, for the improved sampler, lies below 1.1 for most parameters, with a few exceptions. On the other hand, for standard Gibbs, the PSRF of most parameters exceed 1.1, indicating poor convergence. Further convergence statistics like effective number of samples and autocorrelation times are reported in the Supplementary Material. Figure 13 shows a section of the sample trace for the cluster mean for both standard and improved Gibbs. Notice the lack of label switching, i.e. the clusters switching places due to the symmetries of the clustering model, for standard Gibbs, which is another sign that the sampler is failing to explore the entirety of the posterior distribution.

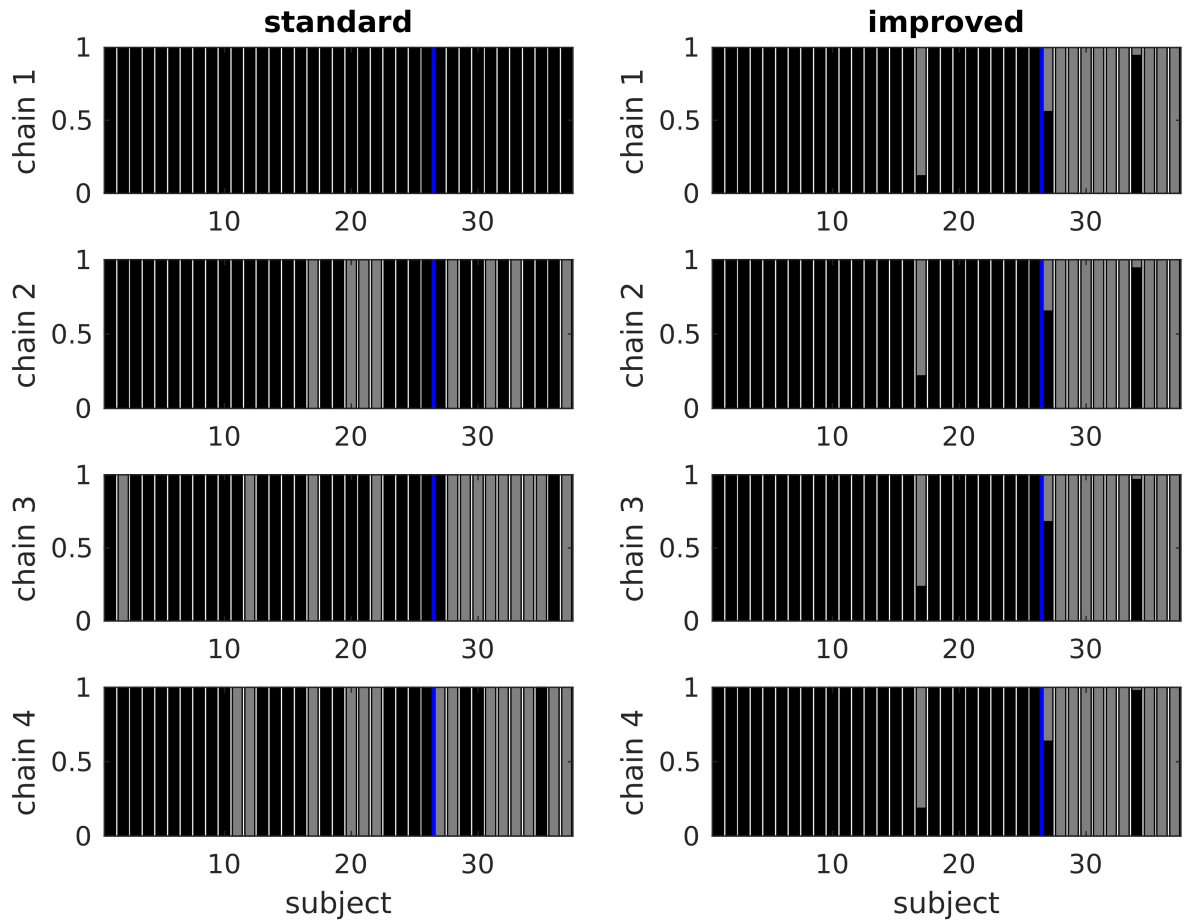

*Figure 10: Subject assignment for experiment 2, estimated for each chain individually for standard (left) and improved (right) Gibbs. The blue line separates the controls (first 26 subject) from patients (last 11 subjects).*

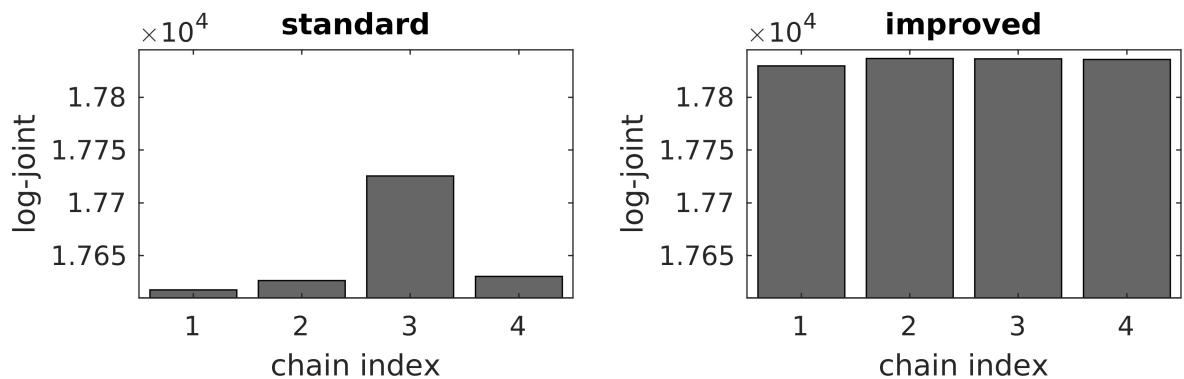

*Figure 11: Mean log-joint probability of individual chains for standard (left) and improved (right) Gibbs sampling.*

In order to compare the performance of our improved Gibbs sampler to HMC on this real-world dataset, we used stan to invert HUGE for this dataset. Sampling 4 chains in parallel, each 200 samples long, with HMC took about $432 \times 10^3$ seconds (approx. 120 hours or 5 days). However, the posterior subject assignments and the PSRF shown in Figure 12 bottom, indicate that a chain length of 200 samples was not sufficient for convergence. Unfortunately, we were not able to repeat the experiment with longer chains, due to limitations on computation time on the HPC cluster used for the experiment.

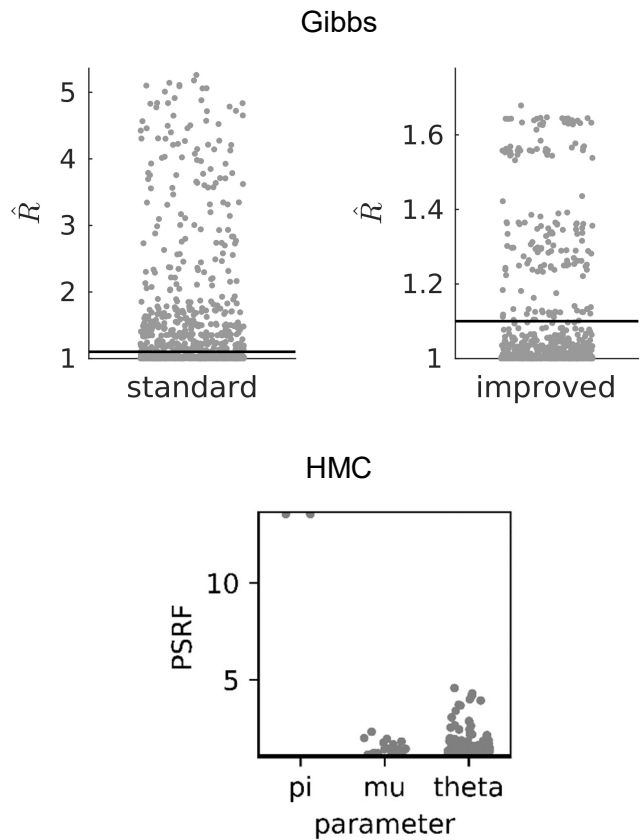

*Figure 12: Top: Range of PSRF values for standard (left) and improved (right) Gibbs for experiment 2. Bottom: PSRF values for HMC with chain length of 200 for different parameter groups.*

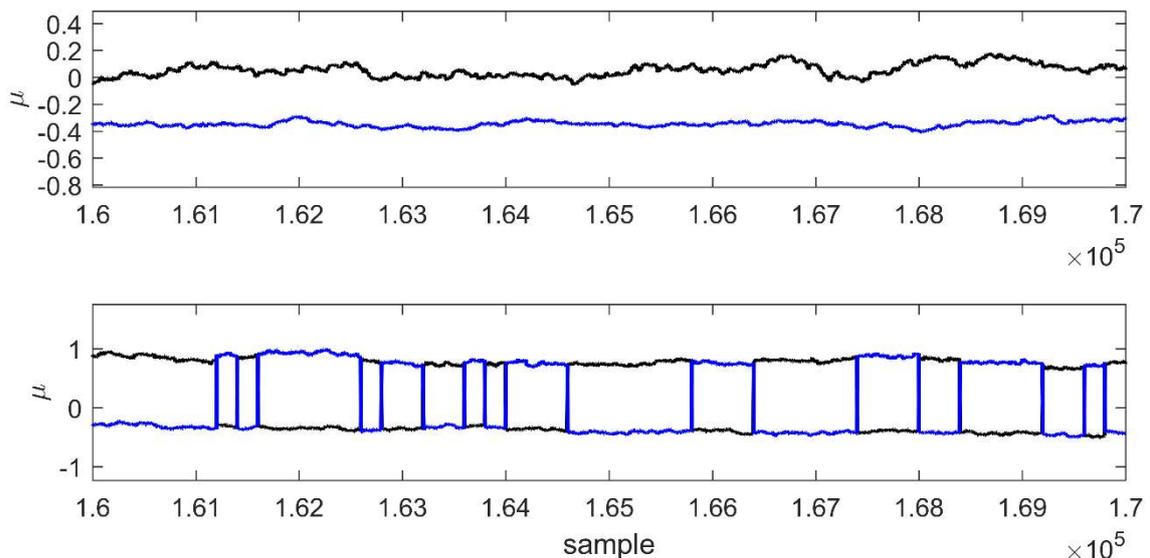

*Figure 13: Sample trace for the cluster mean for standard (top) and improved (bottom) Gibbs.*

It should be noted that the experimental dataset from this section has been analysed in Yao et al. (2018) using VB, which achieved a balanced purity of 91%. While this may seem to indicate that VB is more accurate than MCMC, it is important to note that, unlike for synthetic datasets, the known sub-groups of controls and patients in this experimental dataset only represent an external reference, instead of ground truth parameter settings of the model. In addition, a close inspection of the results reveals that the difference in balanced accuracy is due to a single subject (subject 27), who is assigned to the correct group by VB with high certainty, while the MCMC assignment for this subject is ambiguous, which may simply reflect the well-known tendency of VB to return overconfident results.

## 4 Discussion

In this technical note, we introduced a set of proposal densities tailored to improving the convergence of MCMC samplers for hierarchical clustering. Comparing our approach to HMC in terms of computational complexity, the analysis in section 2.4 showed no clear theoretical advantage for either method. In practice, computation time may vary depending on the particular application. Hence, we conducted extensive empirical tests on synthetic and real-world datasets, from which several conclusions can be drawn.

First, combining hierarchical clustering with dynamic system models such as DCM presents a formidable challenge to standard MCMC samplers due to the strong posterior correlation present, not only between clustering and DCM parameters, but also among the DCM parameters themselves. In addition, the inherent symmetries of clustering models induce multiple modes in the posterior.

Second, designing specialised proposal densities tailored to the specific challenges posed by the hierarchical clustering model represents an effective solution in practice, leading to better clustering performance in terms of balanced purity and also faster convergence. At the same time, the proposal densities we introduced in section 2.2 have only a single free parameter (i.e., the mixing ratio between proposal distributions) which would need to be tuned for optimal performance.

Third, our experiments revealed that even state-of-the-art general purpose Monte Carlo methods, such as HMC, which were specifically designed to avoid random walk behaviour and efficiently sample from highly complex target distributions, struggle to converge reliably in timeframes as often required for solving computational problems (i.e. days). Despite being more efficient, i.e. requiring fewer samples to explore the target distribution, HMC takes longer to run than Metropolized Gibbs sampling with our special proposal distributions and, for our empirical data, did not converge over the timeframe available on our local shared cluster (5 days). This indicates that, in practice, the overhead of having to simulate the Hamiltonian dynamics negate the advantage afforded by HMC of being able to obtain more independent samples.

Concerning the last point, it is important to note that we do not believe this observation to be a consequence of inefficient implementation, since we used the HMC implementation provided by stan, which translates the model into a C program and compiles it before running the sampler. This means that, in our experiments, the HMC sampler enjoyed the run time advantages of a compiled language, while our Gibbs sampler was implemented in Matlab, with only the numerical integration of the DCM evolution equation being implemented in C. For both methods, individual chains were run in parallel.

Fourth, we would like to note that, although we have tested our improved Gibbs sampler on HUGE, which is a DCM for fMRI-based hierarchical clustering model, our proposed approach can be extended easily to hierarchical clustering based on other generative models or other

DCM variants, such as DCM for EEG. Clearly, in contrast to the HMC sampler in stan (and samplers provided by other toolboxes), our approach does not generalise across models; instead, the proposal densities need to be adapted to the specific model used. However, as demonstrated above, an analysis of the dependency structure in the overall hierarchical clustering model provides guidance for this relatively straightforward step.

Clustering subjects of heterogeneous populations is finding increasing application in neuroimaging, particularly in application to psychiatry (Brodersen et al., 2014; Dinga et al., 2019; Drysdale et al., 2017; Feczko et al., 2018; Feczko et al., 2019; Marquand, Wolfers, Mennes, Buitelaar, & Beckmann, 2016; Wolfers et al., 2019), where the considerable heterogeneity of diseases according to current classifications represents a central problem (Stephan et al., 2016). The ability to detect unknown subgroups could greatly improve our means of stratifying psychiatric populations and conduct more powerful clinical trials. This is particularly the case when subgroups are not simply defined in terms of data features, but are interpretable in terms of physiological processes that generated the data – a main motivation behind hierarchical formulations of GE. The technical improvement proposed in this paper may find useful application in future studies that utilise GE for a stratification of heterogeneous disorders.

# 5 Acknowledgement

The work presented in this paper has been funded by the René und Susanne Braginsky Foundation and the University of Zurich (to KES). We would like to thank Prof. Alex Leff for his generous permission to use the fMRI speech perception dataset.

The authors report no conflicts of interest.